\documentclass[final,3p,times,twocolumn]{elsarticle}

\usepackage{graphicx}

\usepackage{amssymb}

\journal{Astroparticle Physics}

\begin{document}

\begin{frontmatter}


\title{Self-shielding effect of a single phase liquid xenon detector\\
for direct dark matter search}

\author[label1]{A. Minamino\corref{cor1}\fnref{fnlabel1}}
\ead{minamino@suketto.icrr.u-tokyo.ac.jp}
\author[label1]{K. Abe}
\author[label1]{Y. Ashie}
\author[label1]{J. Hosaka}
\author[label1]{K. Ishihara}
\author[label1]{K. Kobayashi}
\author[label1]{Y. Koshio}
\author[label1]{C. Mitsuda\fnref{fnlabel2}}
\author[label1]{\\S. Moriyama}
\author[label1,label3]{M. Nakahata}
\author[label1]{Y. Nakajima}
\author[label1]{T. Namba\fnref{fnlabel3}}
\author[label1]{H. Ogawa}
\author[label1]{H. Sekiya}
\author[label1]{M. Shiozawa}
\author[label1,label3]{Y. Suzuki}
\author[label1]{\\A. Takeda}
\author[label1]{Y. Takeuchi}
\author[label1]{K. Taki}
\author[label1]{K. Ueshima}

\author[label4]{Y. Ebizuka}
\author[label4]{A. Ota}
\author[label4]{S. Suzuki}

\author[label5]{H. Hagiwara}
\author[label5]{Y. Hashimoto}
\author[label5]{\\S. Kamada}
\author[label5]{M. Kikuchi}
\author[label5]{N. Kobayashi}
\author[label5]{T. Nagase}
\author[label5]{S. Nakamura}
\author[label5]{K. Tomita}
\author[label5]{Y. Uchida}

\author[label6]{Y. Fukuda}
\author[label6]{T. Sato}

\author[label7]{\\K. Nishijima}
\author[label7]{T. Maruyama}
\author[label7]{D. Motoki}

\author[label8]{Y. Itow}

\author[label12]{Y. D. Kim}
\author[label12]{J. I. Lee}
\author[label12]{S. H. Moon}

\author[label13]{K. E. Lim\fnref{fnlabel4}}

\author[label14]{J.P Cravens\fnref{fnlabel5}}
\author[label14]{\\M.B. Smy\fnref{fnlabel6}}
\author{\\the XMASS Collaboration}

\address[label1]{Kamioka Observatory, Institute for Cosmic Ray Research,
The University of Tokyo, Kamioka, Hida, Gifu 506-1205, Japan}
\address[label3]{Institute for the Physics and Mathematics of the Universe,
University of Tokyo, Kashiwa, Chiba 277-8582, Japan}
\address[label4]{Faculty of Science and Engineering,
Waseda University, Shinjyuku-ku, Tokyo 162-8555, Japan}
\address[label5]{Department of Physics, Faculty of Engineering,
Yokohama National University, Yokohama, Kanagawa 240-8501, Japan}
\address[label6]{Department of Physics,
Miyagi University of Education, Sendai, Miyagi 980-0845, Japan}
\address[label7]{Department of Physics,
Tokai University, Hiratsuka, Kanagawa 259-1292, Japan}
\address[label8]{Solar Terrestrial Environment Laboratory,
Nagoya University, Nagoya, Aichi 464-8602, Japan}
\address[label12]{Department of Physics,
Sejong University, Seoul 143-747, Korea}
\address[label13]{Department of Physics,
Ewha W. University, Seoul 120-750, Korea}
\address[label14]{Department of Physics and Astronomy,
University of California, Irvine, Irvine, CA 92697-4575, USA}

\cortext[cor1]{Corresponding author, Tel./Fax: +81-578-85-9635,}
\fntext[fnlabel1]{Present Address: High Energy Physics Group, 
Department of Physics, Faculty of Science, Kyoto University, 
Kitashirakawa, Oiwake-cho, Sakyo-ku, Kyoto,
606-8502, Japan}
\fntext[fnlabel2]{Present Address: 
Accelerator Division, 
Japan Synchrotron Radiation Research Institute (JASRI), 
Kouto, Sayo-cho, Sayo-gun,
Hyogo,
679-5198, Japan}
\fntext[fnlabel3]{Present Address: 
International Centre for Elementary Particle Physics, 
The University of Tokyo, 
Hongo, Bunkyo-ku,
Tokyo,
113-0033, Japan}
\fntext[fnlabel4]{Present Address: 
Department of Physics, 
Columbia University, 
538 W. 120th St New York,
NY,10027, USA}
\fntext[fnlabel5]{Present Address: 
Department of Physics, 
University of Texas, 
1 University Satation, Austin,
TX,78712-1081, USA}
\fntext[fnlabel6]{Present Address: 
Institute for the Physics and Mathematics of the Universe,
University of Tokyo, Kashiwa, Chiba 277-8582, Japan}

\begin{abstract}
Liquid xenon is a suitable material for a dark matter search.
For future large scale experiments, single phase detectors are attractive due to their simple configuration and scalability.
However, in order to reduce backgrounds, they need to fully rely on liquid xenon's self-shielding property.
A prototype detector was developed at Kamioka Observatory to establish vertex and energy reconstruction methods and to demonstrate the self-shielding power against gamma rays from outside of the detector.
Sufficient self-shielding power for future experiments was obtained.
\end{abstract}


\end{frontmatter}



\section{Introduction}
\label{sec_intro}
Recent results from cosmic microwave background (CMB), large scale structure (LSS) and type Ia supernovae observations have yielded a standard model of cosmology:
a flat universe consisting of more than 70\% dark energy and about 23\% dark matter with the remainder ordinary (baryonic) matter \cite{schmidt,spergel,tegmark}.
The widely discussed (nonbaryonic) dark matter candidate is Weakly Interacting Massive Particles (WIMPs).
The newest and lightest particle, the neutralino (predicted by Supersymmetry (SUSY)), would be an ideal WIMP candidate \cite{jungman}.

WIMP dark matter can elastically scatter off the target nuclei of a terrestrial detector, transferring energy to the nuclei.
Dark matter can be observed by detecting this released interaction energy.
But the main challenge is to identify a WIMP recoil event over many background events.
Recent best limits on an upper bound for the cross section of WIMP spin independent interactions are given by the CDMSII \cite{cdms2}, XENON10 \cite{xenon10} and ZEPLIN-III \cite{zeplin3} experiments.
To increase the sensitivity,
a detector with larger mass and low background is required.

XMASS was originally proposed as a 10-ton multi-purpose single phase liquid xenon detector aiming to detect pp-solar neutrinos, dark matter, and double beta decay \cite{xmass-origin}.
For a large scale experiment, single phase detectors are attractive due to their simple detector configuration and scalability.
We have adopted a step by step approach and will initially develop an 800kg-sized detector ($\sim$100kg fiducial mass) dedicated to a dark matter search.
Liquid xenon has a large light yield comparable to that of NaI(Tl) therefore allowing us to set a low energy threshold and achieve good acceptance for dark matter.
Because of liquid xenon's large atomic number (Z = 54) and high density ($\sim$ 3 g/cm$^{3}$), background $\gamma$-rays are quickly attenuated near the outer edge of the liquid xenon volume which makes a low background volume around the center of the detector.
This self-shielding of the environmental $gamma$-ray background is the key concept of the detector.

In this paper, we describe the results from a prototype detector by which we have measured basic properties of a single phase liquid xenon detector.
We have established vertex and energy reconstruction methods and demonstrated the self-shielding power for background reduction.

\section{Experimental set-up}
\label{sec_setup}

\subsection{Environmental backgrounds}
The WIMPs signal can be identified over many background events.
For nuclear recoil events, the largest background contribution originates from neutrons.
Kamioka Underground Laboratory (the site for XMASS) is at a depth of 2700 m.w.e. (meters water equivalent), and has a smaller neutron flux than a surface laboratory by more than two orders of magnitude \cite{ydkim_lrt2006}.
The muon flux is also reduced by about five orders of magnitude \cite{sk_muon} making the cosmogenic activation of xenon negligible.
During the future operation of the 800kg detector, environmental neutrons and gamma-rays will be reduced to negligible levels by a water shield of more than 2m thickness.
However, backgrounds from PMTs and other materials used for the detector and supporting structures will not be reduced by such means mentioned above.
Self-shielding is the most effective suppressor for those backgrounds.

\subsection{Liquid xenon scintillator}
The physical properties of liquid xenon are summarized in Table \ref{tab:physical_properties_lxe}.
\begin{table}[h]
\begin{center}
\vskip0.3cm
\begin{tabular}{|c|c|c|}
\hline
Parameter & Value \\
\hline
\hline
Atomic number & 54 \\
\hline
Mass number & 131.29 \cite{handbook_chem_phys} \\
\hline
Boiling point at 1 atm & 165.1 K \cite{handbook_chem_phys} \\
\hline
Melting point at 1 atm & 161.4 K \cite{handbook_chem_phys} \\
\hline
Density at 161.5 K & 2.96 g/cm$^{3}$ \cite{sinnock} \\
\hline
Radiation length & 28.7 mm \cite{PDG} \\
\hline
Peak emission wavelength & 178 nm \cite{jortner,schwenter} \\
\hline
Spectral width (FWHM) & $\sim$14nm \cite{jortner,schwenter} \\
\hline
Average energy required & 23.7$\pm$2.4 eV \cite{doke_light_yield}\\
for one photon production   & 14.2 eV \cite{seguinot}\\
\hline
Scint. Absorption length & $\ge$100cm \cite{meg}\\
\hline
Rayleigh scattering length & 60 cm \cite{rayleigh} \\
\hline
Refractive index & 1.61$\pm$0.1 \cite{nakamura_refractive} \\
\hline
\end{tabular}
\end{center}
\caption{Physical properties of liquid xenon}
\label{tab:physical_properties_lxe}
\end{table}

Existing reports vary on the average energy required to produce one photon by ionizing radiation.
Also, the absorption length changes with the amount of water contamination in liquid xenon.
We have checked these properties by using the prototype detector in order to demonstrate the feasibility of achieving the target sensitivity for a WIMP search with a large scale single phase detector.

The relative scintillation efficiency of nuclear recoils to electron recoils is called the quenching factor, $f_{Q}$.
The nuclear recoil energy, $E_{R}$, is converted into electron-equivalent visible energy, $E_{V}$, as follows:
\begin{eqnarray}
E_{V} = f_{Q}E_{R}.
\end{eqnarray}
The unit of visible energy is keV electron equivalent (keVee),
and the quenching factor of liquid xenon has been measured by several groups \cite{quenching1, quenching2, quenching3}.

\subsection{Detector}
The detector is a cubic vessel of 30cm in height made from oxygen free high conductivity (OFHC) copper with an inner volume of 27 litters.

High purity 5N ($>99.999\%$) aluminum and 4N ($>99.99\%$) MgF$_{2}$
were evaporated on the inner surface as a reflective mirror
for the scintillation light.

\begin{figure}[h]
\begin{center}
\includegraphics[scale = 0.33]{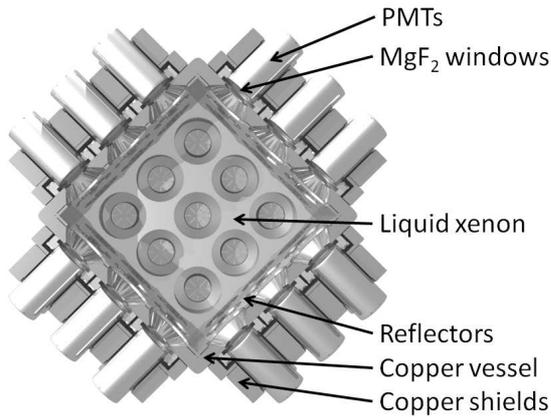}
\caption{Schematic view of the prototype detector. The cubic vessel containing liquid xenon is made from OFHC copper with an inner volume of 27 litters.
Scintillation photons are detected by PMTs through MgF$_{2}$ windows.
}
\label{fig:prototype_fig}
\end{center}
\end{figure}

\begin{figure}[h]
\begin{center}
\includegraphics[scale = 0.38]{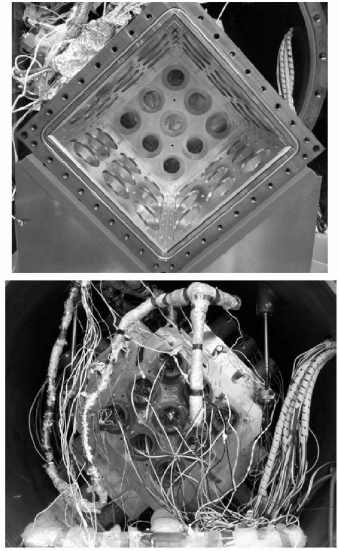}
\caption{Pictures of the prototype detector.
Reflectors are attached on the inner wall of the cubic vessel (top).
Nine PMTs are attached on each side of the vessel and detect the  scintillation photons (bottom).
}
\label{fig:prototype_pic}
\end{center}
\end{figure}

Nine PMTs are attached to each exterior side of the vessel, and scintillation photons are detected by 54 PMTs through MgF$_{2}$ windows.
Photo-coverage of the detector is about 16 \%.
Gaps between the PMTs and the MgF$_{2}$ windows were filled with a high refractive index material, ``Krytox\textsuperscript{\textregistered} 16350,'' manufactured by DuPont (n = 1.35 for 178 nm photons), allowing us to successfully increase the photoelectron yield of the detector by reducing total reflection between the MgF$_{2}$ windows and vacuum.
A schematic view and pictures of the prototype detector are shown in Figures \ref{fig:prototype_fig} and \ref{fig:prototype_pic}.

\subsection{Low background PMTs}
We have developed a low background 2-inch PMT, R8778ASSY, with Hamamatsu Photonics K.K.
The PMT has reasonably high quantum efficiency, $\sim$30 \%, for 178 nm photons and can be operated at 173 K.
Components of the PMT were examined individually and selected to reduce radioactive backgrounds.
Table \ref{tab_ri_r8778} shows the contamination of the radioactive impurities measured by a high purity Ge (HPGe) detector at Kamioka.

\begin{table}[h]
 \begin{center}
 \begin{tabular}{|c|c|}
 \hline
 U (mBq/PMT) &  $18\pm2$ \\
 \hline
 Th (mBq/PMT) & $6.9\pm1.3$ \\
 \hline
 K (mBq/PMT) & $140\pm20$ \\
 \hline
 $^{60}$Co (mBq/PMT) & $5.5\pm0.9$ \\
 \hline
 \end{tabular}
 \end{center}
 \caption{Radioactive impurities in the R8778ASSY PMT measured by a HPGe detector \cite{nakahata_dark2007}.
The uranium and thorium chains were assumed to be in radioactive equilibrium.}
 \label{tab_ri_r8778}
\end{table}

\subsection{Radiation shields}
\label{sec_radiation_shields}
As the expected count rate caused by WIMPs is extremely low, radiation shields are inevitable for dark matter search experiments.
A set of shields as listed in Table \ref{tab:shields_prototype} was adopted.
\begin{table}[h]
 \begin{center}
 \begin{tabular}{|c|c|}
 \multicolumn{2}{c}{Outer shields}\\
 \hline
 material  &  thickness\\
 \hline
 \hline
 polyethylene & 15 cm\\
 \hline
 boric acid & 5 cm\\
 \hline
 lead & 15 cm\\
 \hline
 EVOH sheet & - \\
 \hline
 OFHC copper & 5 cm\\
 \hline
 \multicolumn{2}{c}{}\\
 \multicolumn{2}{c}{Inner shields}\\
 \hline
 Inner OFHC copper shield & 6.7 cm\\
 \hline
 \end{tabular}
 \end{center}
 \caption{Material of the radiation shields. Listed from outside-in.}
 \label{tab:shields_prototype}
\end{table}

The polyethylene shield and the boric acid powder shield act as a neutron moderator and absorber, respectively.
The lead acts as the shield against environmental $\gamma$-rays.
The Ethylene Vinyl Alcohol (EVOH) sheets with a reduced radon air purge system reduces radon gas from the inside of the shield.
The OFHC copper shield acts as a shield against bremsstrahlung $\gamma$-rays and X rays from the lead shield and is closest to the PMTs.

\subsection{Operation}
The detector and xenon gas line were evacuated and baked for two weeks at a temperature of about 373 K.
The de-gas rate inside the detector and the gas line (total volume $\sim$ 57 liters) after baking was on the order of $10^{-5}$ Pa/second at room temperature. 
Xenon gas was passed through a SAES getter (Model PS4-MT3-R-1) before filling the detector to remove impurities which shorten the attenuation length in liquid xenon.
The radioactive impurity krypton cannot be removed with the getter so it was removed by a distillation tower before operation \cite{nakahata}.

A Gifford-McMahon (GM) refrigerator was used to liquefy the xenon.
It also kept the detector at 173 K after filling with liquid xenon.
For thermal insulation, the detector was suspended inside a vacuum chamber.
The pressure inside the vacuum chamber varied from 10$^{-1}$ to 10$^{-3}$ Pa depending upon the temperature of the detector (room temperature to 173 K).

\subsection{Electronics and data acquisition system}
The output signal from each PMT was fed into ADCs (V792, CAEN), a 250MHz FADC (STR7515, BASTIAN Technology), and leading edge discriminators (V814, CAEN).
The gain of the 54 PMTs was set to be 8.25$\times$10$^{6}$.
The threshold level of the discriminator was 0.25 photoelectrons, and the timing of each PMT was recorded by TDC modules (Tristan/KEK Online, TKO standard).
The detector trigger required a coincidence of at least three hits within 95 ns.
Trigger timing and veto timing were recorded by a TRG(TRiGger) module (VME standard) \cite{trg}.
The live time of the measurement was calculated from the TRG information.

\section{Detector Simulation}
\label{sec_simulation}
A detector simulator program was developed by using the framework of the GEANT3 package \cite{geant3}.
In the simulation code, tracks of particles, scintillation processes, propagation of scintillation photons, and the response of PMTs are simulated.

In connection with the propagation of charged particles, scintillation photons are generated according to a linear energy transfer (LET) dependence of the scintillation yield \cite{doke_let} given by the equation:
\begin{eqnarray}
dL/dE = \left[ A(dE/dx) / (1 + B(dE/dx)) \right] + \eta_{0},
\label{eq:let_dependence}
\end{eqnarray}
where $dL/dE$ is normalized to 1 in the limiting case of $dE/dx \rightarrow \infty$.
A, B, and $\eta _{0}$ are adjustable parameters, where $\eta _{0}$ is the scintillation yield at zero electric field in the limit of zero LET.
In this paper, the values that manifest the best fit for liquid xenon \cite{doke_let} are used as $\eta _{0}$, A, and B.
\begin{eqnarray}
&&\eta_{0} = 0 \label{eq:par_eta0} \\
&&A = B = 1.22. \label{eq:par_a_b}
\end{eqnarray}

For the propagation of scintillation photons in liquid xenon, Rayleigh scattering and absorption are considered in our simulation code.
At the boundary of liquid xenon and other transparent materials, scintillation photons are refracted or reflected according to the Fresnel equations.
Light reflection and absorption in detector materials, such as the surface of the PMTs and the mirror on the inner wall of the detector, are also simulated.
The light yield and absorption length of liquid xenon as well as the reflectance at the mirror on the inner wall were tuned to reproduce $\gamma$-ray calibration data as discussed in Section\ref{absorption_length_light_yield}.

For the simulation of the PMT output of a single photoelectron, we employed an observed one-photoelectron distribution as a probability distribution function of the PMT output.
For the simulation of multiple photoelectron signals, the PMT output was derived by summing up the PMT outputs for single photoelectrons with the required number of times.

\section{Vertex and energy reconstruction}
\label{sec_reconstruction}
As discussed in Section \ref{sec_intro}, the self-shielding of liquid xenon for $\gamma$-ray backgrounds is the key idea of the detector.
Therefore the vertex reconstruction is a crucial issue for the detector.

The vertex and energy of each event in the prototype detector is reconstructed with the light yield distribution of the 54 PMTs.
Light yield patterns of the PMTs from scintillation light at a single vertex are estimated by the detector simulation in a 30 cm cubic lattice with steps every 3 mm (1030301 points).
The vertex is reconstructed with the following likelihood of a Poisson distribution by comparing the detected light yield distribution with the simulated ones:
\begin{eqnarray}
\log (L)= \sum ^{54}_{i=1} \ \log \left( \frac{\exp ^{-\mu _{i}} \mu _{i} ^{n}}
                                    {n!} \right).  \label{eq:likelihood_recon}
\end{eqnarray}
Here, $n$ is the observed number of photoelectrons and $\mu _{i}$ is
\begin{eqnarray}
\mu _{i} = F_{i}(x,y,z) \times \textrm{num. of generated photons},
\label{eq:mu_recon}
\end{eqnarray}
where $F_{i}(x,y,z)$ is the probability of observing one photoelectron at PMT No. $i$ when a single photon is generated at vertex $(x,y,z)$. ``num. of generated photons'' is calculated as
\begin{eqnarray}
\textrm{num. of generated photons} = \frac{\sum ^{54} _{i=1} pe(i)}{\sum ^{54} _{i=1} F_{i}},
\end{eqnarray}
where $pe(i)$ is the light yield detected by PMT No. {\it{i}}.

The vertex with maximum likelihood is selected as the reconstructed vertex.
The reconstructed energy is then calculated as
\begin{eqnarray}
E\ \textrm{[keV]} =\frac{\textrm{``num. of generated photons''}}
        {N_{photon}}.
\end{eqnarray}
Here, $N_{photon}$ is the number of generated photons per 1 keV measured with 662 keV $\gamma$-rays from a $^{137}$Cs source.

\section{Results}
\label{sec_results}
\begin{figure}[h]
\begin{center}
\includegraphics[scale = 0.38]{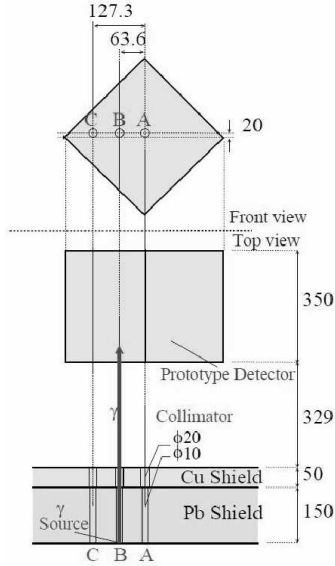}
\caption{Relative locations of the $\gamma$-ray collimators (A, B, C) and the prototype detector.}
\label{fig:hole_abc}
\end{center}
\end{figure}

$\gamma$-rays were used to test various aspects of the detector.
These $\gamma$-rays are collimated through the radiation shield as shown in Figure \ref{fig:hole_abc}.

\subsection{Measurement of xenon properties} \label{absorption_length_light_yield}
Characteristics of liquid xenon
(e.g. absorption length and scintillation light yield)
and the reflectance of the inside wall of the detector
were measured by the collimated $\gamma$-rays from $^{137}$Cs and $^{60}$Co.
Here, the Rayleigh scattering length of liquid xenon for 178 nm photons
was set to be 60 cm \cite{rayleigh}.
Light yield distributions were compared with
those of the simulation by using the $\chi ^{2}$ method,
and the best fit parameter set with systematic errors were obtained as follows:
\begin{itemize}
\item Absorption length = 66$\pm$10 cm
\item Reflectance of the mirror = 52+4-6 \%
\item Average energy required for one photon production 
= 17.8$\pm$0.9 eV 
(obtained for 662 keV $\gamma$-rays)
\end{itemize}
These parameters are used for the following analysis.

The average energy required for one photon production
obtained for 662 keV $\gamma$-rays
is smaller than that estimated for 1 MeV electrons
by T. Doke et al. \cite{doke_light_yield}
and is larger than that obtained for an electron beam with kinetic energy
$<$ 100 keV by J. S\'eguinot et al. \cite{seguinot}.

\subsection{Event reconstruction}
\begin{figure}[h]
\begin{center}
\includegraphics[scale = 0.30]{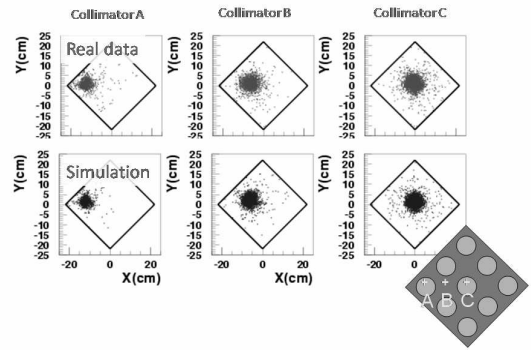}
\caption{The X-Y projection of the reconstructed vertices
of $^{137}$Cs $\gamma$-rays from Collimator A, B and C.
The injection positions of the $\gamma$-rays are shown in the bottom-right drawing.
}
\label{fig:vertex_reconstruction}
\end{center}
\end{figure}

\begin{figure}[h]
\begin{center}
\includegraphics[scale = 0.38]{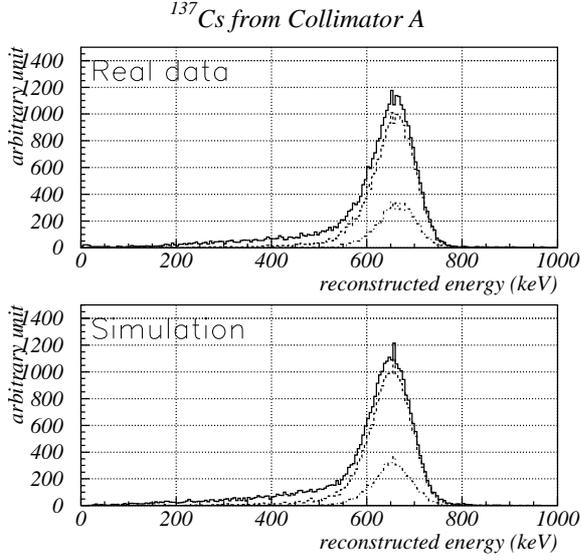}
\caption{Reconstructed energy spectrum
of $^{137}$Cs $\gamma$-rays from Collimator A.
Events reconstructed in full volume, 
20 cm and 10 cm cubes from the center of the detector
correspond to solid lines, dashed lines and broken lines. 
Photoelectric absorption peaks (662 keV) in each fiducial volume
are fit with an asymmetric gaussian.
The energy resolution in each fiducial volume is shown
in Table \ref{table:e_spectrum_cs137a_fvcut}.
}
\label{fig:e_spectrum_cs137a_fvcut_real_mc}
\end{center}
\end{figure}

The vertex and energy reconstruction methods
were evaluated by using the $\gamma$-ray data
($^{137}$Cs at 662 keV) from Collimator A, B and C.
The X-Y projections of the reconstructed vertices of real data
cluster around the $\gamma$-ray injection points
and agree well with that of the detector simulation
as shown in Figure \ref{fig:vertex_reconstruction}.
This demonstrates that the vertex reconstruction works as expected.
The reconstructed energy spectrum of real data
in successive fiducial volumes
(full volume, 20 cm and 10 cubics from the center of the detector)
agree well with that of the detector simulation 
as shown in Figure \ref{fig:e_spectrum_cs137a_fvcut_real_mc}
and Table \ref{table:e_spectrum_cs137a_fvcut}.
Hence the energy reconstruction also works as expected.

\begin{table}[h]
\begin{center}
\vskip0.3cm
\begin{tabular}{|c|c|c|}
\hline
  &  mean (keV)  & resolution (\%)  \\
\hline
\hline
\multicolumn{3}{|c|}{Real data} \\
\hline
Full vol. & $670.9\pm0.7$ & $5.57\pm0.09$ \\
20cm cubic     & $672.4\pm1.1$ & $5.47\pm0.13$ \\
10cm cubic     & $676.1\pm1.8$ & $5.20\pm0.21$ \\
\hline
\hline
\multicolumn{3}{|c|}{Simulation} \\
\hline
Full vol. & $662.0\pm0.3$ & $5.83\pm0.07$ \\
20cm cubic     & $663.4\pm1.1$ & $5.71\pm0.11$ \\
10cm cubic     & $663.8\pm1.9$ & $5.59\pm0.20$ \\
\hline
\end{tabular}
\end{center}
\caption{Energy resolution of the prototype detector
in each fiducial volume 
for 662 keV $\gamma$-rays ($^{137}$Cs from Collimator A).}
\label{table:e_spectrum_cs137a_fvcut}
\end{table}

\subsection{Non-linearity of scintillation yield}
\begin{figure}[h]
\begin{center}
\includegraphics[scale = 0.3]{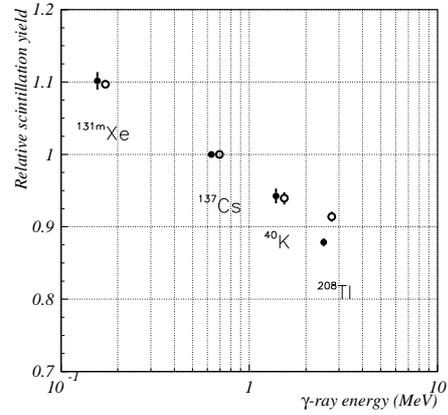}
\caption{Relative scintillation yield
of real data (closed circle) 
and simulation (open circle) for various $\gamma$ rays.
Vertical axis is normalized with 662-keV $\gamma$ rays ($^{137}$Cs).
Both real data and simulation have a non-linearity, 
$\sim \pm 10$ \% from 662 keV.
}
\label{fig:scinti_eff_real_mc}
\end{center}
\end{figure}

\begin{figure}[h]
\begin{center}
\includegraphics[scale = 0.3]{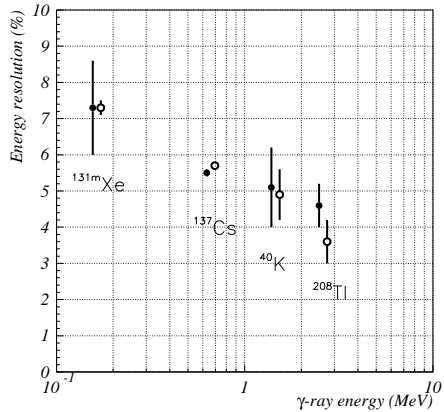}
\caption{Reconstructed energy resolution
of real data (closed circle) 
and simulation (open circle) for various $\gamma$ rays.
}
\label{fig:resolution_real_mc}
\end{center}
\end{figure}

The reconstructed energy spectra of the following $\gamma$-ray sources
are compared with that of the simulation:
\begin{itemize}
\item $^{131m}$Xe uniformly inside the detector (164 keV).
\item $^{137}$Cs source from Collimator A (662 keV).
\item $^{40}$K source inside the radiation shield (1461 keV).
\item $^{208}$Th source inside the radiation shield (2615 keV).
\end{itemize}
Events reconstructed in the 20 cm cubic volume
are used for comparison.
As shown in Figures \ref{fig:scinti_eff_real_mc} and
\ref{fig:resolution_real_mc} and Table \ref{table:eres},
real data have a non-linear scintillation yield and 
energy resolution which cannot be explained by Poisson statistics.
By introducing a LET dependence of the scintillation yield
(discussed in Section \ref{sec_simulation}),
the simulation successfully reproduces the real data well
except for $^{208}$Th.
To reproduce the $^{208}$Th data with the simulation,
the scintillation yield must have a larger LET dependence
than one expressed in Equation (\ref{eq:let_dependence}).

\begin{table}[h]
\begin{center}
\vskip0.3cm
\begin{tabular}{|c|c|c|c|}
\hline
 \multicolumn{2}{|c|}{}  & mean (keV) & $\sigma$ (\%) \\
\hline
 $^{131m}$Xe   & Real data  & $183.5\pm2.0$ & $7.3\pm1.3$ \\
 164 keV      & Simulation & $180.3\pm0.2$ & $7.3\pm0.2$ \\
\hline
 $^{137}$Cs    & Real data  & $672.4\pm1.1$ & $5.5\pm0.1$ \\
 662 keV      & Simulation & $663.4\pm1.1$ & $5.7\pm0.1$ \\
\hline
 $^{40}$K      & Real data  & $1399\pm15$ & $5.1\pm1.1$ \\
 1461 keV     & Simulation & $1376\pm12$ & $4.9\pm0.7$ \\
\hline
 $^{208}$Tl    & Real data  & $2334\pm15$ & $4.6\pm0.6$ \\
 2615 keV     & Simulation & $2395\pm18$ & $3.6\pm0.6$ \\
\hline
\end{tabular}
\end{center}
\caption{Fitted results of the photoelectric absorption peaks of
various $\gamma$-ray sources.
Peaks are fitted with asymmetric gaussians.
}
\label{table:eres}
\end{table}

\subsection{Self-shielding power of liquid xenon}
\begin{figure}[h]
\begin{center}
\begin{minipage}{0.47\textwidth}
\begin{center}
\includegraphics[scale = 0.3]{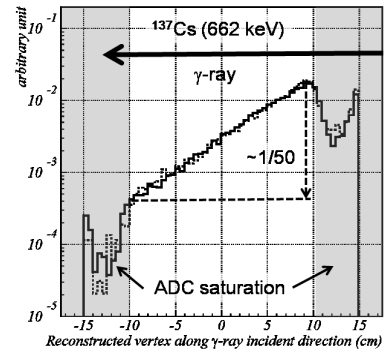}
\end{center}
\end{minipage}
\begin{minipage}{0.47\textwidth}
\begin{center}
\includegraphics[scale = 0.3]{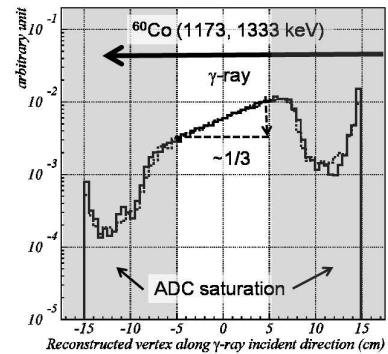}
\end{center}
\end{minipage}
\caption{Reconstructed vertex along the particle incident direction
for 662 keV ($^{137}$Cs) and 1173, 1333 keV ($^{60}$Co) $\gamma$ rays.
The distributions of the real data (solid line) agree well with 
that of the detector simulation (dotted line).
Both ends of the distributions are distorted by cutting ADC saturated events.
}
\label{fig:self_shielding_cs_co_hole_a}
\end{center}
\end{figure}

The self-shielding power of liquid xenon was evaluated with
$\gamma$-ray data, $^{137}$Cs (662 keV) and 
$^{60}$Co (1173 and 1333 keV) from Collimator A.
If a $\gamma$-ray vertex is near a PMT,
it is difficult to accurately reconstruct the vertex 
because the ADC of the PMT saturates at 200 photoelectrons.
Therefore these ADC saturated events were removed.
The reconstructed vertices along the $\gamma$-ray incident direction 
(reconstructed $z$) was checked for events 
where the reconstructed vertices were within 10 cm from the center
on the $xy$ plane as shown in Figure \ref{fig:self_shielding_cs_co_hole_a}.
Both ends of the distributions are distorted by cutting
ADC saturated events.
The number of events decreases as a function of the distance from the injected position
of the $\gamma$-rays which is expected from self-shielding.
662 keV $\gamma$-rays from $^{137}$Cs are attenuated by a factor of about 50 
over a distance of 20 cm.
While 1173 and 1333 keV $\gamma$-rays from $^{60}$Co are 
attenuated by a factor of about 3
over a distance of 10 cm.
The distributions of real data agree well with
that of the detector simulation.

\section{Conclusion} \label{sec_conclusion}
A prototype single phase liquid xenon detector was developed.
We have tested the performance of a single phase liquid xenon detector,
have measured the physical properties of liquid xenon,
and have tested the vertex and energy reconstruction methods.
By introducing a LET dependence in the scintillation yield,
the simulation successfully reproduces
the non-linear scintillation yield ($\pm$ 10 \%) and 
the energy resolution of real data between 164 keV and 2615 keV.
We have demonstrated that the self-shielding of liquid xenon 
for background $\gamma$-rays is quite effective.

The 800kg single phase liquid xenon detector is optimized 
for a dark matter search.
The photo coverage is $\sim$ 70 \% and the expected photoelectron yield
is $\sim$ 5 p.e./keVee for an event at the detector center.
This large photoelectron yield enables us to set the energy threshold
to 5 keVee.
With a fiducial volume cut using the reconstructed vertex,
an ultra low background level as low as $10^{-4} /kg/day/keV$ 
can be achieved in the center 40 cm diameter fiducial volume.
The expected sensitivity of the detector for dark matter
is about 2 orders magnitude better than the present known limit
\cite{cdms2}
assuming 5 years of data with 100 kg fiducial volume.

\section{Acknowledgements}
We gratefully acknowledge the cooperation of Kamioka Mining
and Smelting Company.
This work was supported by Grant-in-Aid for Scientific Research on Priority Areas.
We are supported by Japan Society for the Promotion of Science.


\end{document}